\documentclass[11pt]{article}

\title{Automated Protein Structure Classification:\\ A Survey}
\author{
Oktie Hassanzadeh
\\ \tt oktie@cs.toronto.edu
}

\date{January 2008}

\begin{document}

\maketitle

\begin{abstract}
 Classification of proteins based on their structure provides a valuable resource for studying protein structure, function and evolutionary relationships. With the rapidly increasing number of known protein structures, manual and semi-automatic classification is becoming ever more difficult and prohibitively slow. Therefore, there is a growing need for automated, accurate and efficient classification methods to generate classification databases or increase the speed and accuracy of semi-automatic techniques. Recognizing this need, several automated classification methods have been developed. In this survey, we overview recent developments in this area. We classify different methods based on their characteristics and compare their methodology, accuracy and efficiency. We then present a few open problems and explain future directions.
\end{abstract}

\subsection*{Introduction} 
 Classification of protein structures is an interesting and challenging problem in the field of computational biology that plays an important role in several tasks for studying protein function. These tasks include protein structure and function prediction, studying structural and evolutionary relationships between proteins, and identification of potential functional residues and binding sites. Several classification databases exist \cite{FSSP, CATH, SCOP, HOMSTRAD, MMDB, 3Dee} from which SCOP\footnote{Structural Classification Of Proteins} \cite{SCOP} and CATH\footnote{CATH is an acronym of the four main levels in its classification: Class, Architecture, Topology and Homologous superfamily} \cite{CATH} are the most widely used and active databases. These databases are updated intermittently using manual and semi-automatic methods. For example, SCOP has been updated every seven months on average during the last seven years, while CATH has been updated annually. 

On the other hand, the number of newly determined protein structures is constantly increasing. The Protein Data Bank (PDB) \cite{PDB} currently contains 46,051 structures (as of October 2007), of which 6,358 structures were released during the first three quarters of 2007. This is roughly double the number of structures in the year 2004. This rapid increase in the number of structures calls for more efficient, accurate and automated classification methods. 
 Automated methods may not be able to completely replace the manual and semi-automatic databases that incorporate the judgement of an experienced biologist. However, with the rapid  increase in the number of known structures, they can (and currently do) play an important role as a preprocessing step for high-quality manual classification of proteins.
 
 Consequently, several automated classification methods have recently been developed. These methods differ in several aspects including their structure comparison criteria, the type of their input and the output of the classification. In most of the automated classification methods, the goal is to automatically assign a protein structure or domain to an existing class of a manual classification scheme, mainly SCOP and CATH. Some are specifically designed to predict only SCOP or only CATH classes while others provide a more flexible classification framework that is capable of assigning SCOP, CATH or other existing method's classes to input structures. Although the main objective of automated methods is to {\em accurately} classify the input structures, recent methods consider {\em efficiency} as an important criterion in their evaluation as a result of the rapid increase in the number of known structures. Another desirable feature of a classification method is the ability to detect new classes when structures cannot be classified into existing classes.

 In this paper, we present a survey of major existing protein structure classification methods. These methods are listed in Table \ref{tab1} along with their structure comparison criteria. The methods that are purely based on sequence comparison are efficient, but they often fail to identify remote homologs of structurally similar proteins. Methods based on only structure comparison are effective in classifying at fold level, but not necessarily at family and superfamily levels. Methods that combine sequence and structure information for classification are generally more accurate but computationally more expensive.

\begin{table}
\caption{Protein structure classification methods discussed in this paper}
\label{tab1}
\begin{center} 
\begin{tabular}{|rl|c|} \hline
 \multicolumn{2}{|c|}{Method}                                                 & Based on          \\ \hline
 SUPERFAMILY & \footnotesize (Gough et al., 2001 \cite{SUPERFAMILY-Gough01} and 2007 \cite{SUPERFAMILY-Wilson07})& Sequence            \\ \hline
 F2CS (CO)     & \footnotesize (Getz et al., 2002 \cite{byFSSP-Getz02} and 2004 \cite{F2CS-Getz04}) 		& Structure           \\ \hline
 SGM         & \footnotesize (Rogen and Fain, 2003 \cite{SGM-Rogen03}) 		& Structure           \\ \hline
 SCOPmap     & \footnotesize (Cheek et al., 2004 \cite{SCOPMAP-Cheek04})    & Structure/Sequence  \\ \hline
 DTree		 & \footnotesize (\c{C}amoglu et al., 2005 \cite{DTree-Cam05}) 	& Structure/Sequence  \\ \hline
 ProtClass 	 & \footnotesize (Aung and Tan, 2005 \cite{ProtClass-Aung05}) 	& Structure           \\ \hline
 proCC		 & \footnotesize (Kim and Patel, 2006 \cite{proCC-Kim06}) 	    & Secondary Structure \\ \hline
 fastSCOP	 & \footnotesize (Zemla et al., 2007 \cite{STRALCP-Zemla07}) 	& Structure           \\ \hline
\end{tabular}
\end{center}
\end{table}

 Table \ref{tab2} shows the type of the input and the output of the classification methods discussed in this paper. Some methods perform classification on a query protein chain. In these methods, first domain boundaries are identified using an integrated domain prediction technique and then a classification label is assigned to each identified domain. Other methods do not address the domain boundary prediction problem and only classify an input domain. The output of the classification process is a class label or class labels of a classification scheme. Some methods only predict one level of classification in the hierarchy of SCOP or CATH such as SCOP family or superfamily while others are capable of assigning the input structure to all different levels  of the classifications' hierarchy.

\begin{table}
\caption{Input and output in the classification methods}
\label{tab2}
\begin{center} 
\begin{tabular}{|l|c|l|} \hline
 \multicolumn{1}{|c|}{Method}           & Input     & Output \\ \hline
 SUPERFAMILY & Structure & SCOP superfamily \cite{SUPERFAMILY-Gough01} and family \cite{SUPERFAMILY-Wilson07} \\ \hline
 F2CS (CO)   & Domain    & SCOP fold and class \\
             &           & CATH topology and architecture \\ \hline
 SGM         & Domain    & full CATH hierarchy  \\ \hline
 SCOPmap     & Structure & SCOP superfamily and fold  \\ \hline
 DTree		 & Domain    & SCOP family, superfamily and fold  \\ \hline
 ProtClass 	 & Domain    & SCOP fold - possibly other levels  \\ \hline
 proCC		 & Domain    & SCOP family, superfamily and fold  \\
             &           & full CATH hierarchy                \\ \hline
 fastSCOP	 & Structure & SCOP superfamily                   \\ \hline
\end{tabular}
\end{center}
\end{table}

 In the rest of this paper, we briefly overview the classification methodologies of the methods listed in Table \ref{tab1}\footnote{Since these methods differ in many features as shown in Tables \ref{tab1} and \ref {tab2}, we do not group them together based on their features. Instead, they are ordered chronologically by the date of their first published results.}.
 We highlight important features of each method and present discussions on how each of these methods perform comparing with others. We finish this paper by discussing some open problems and several ideas for further improvement in the task of classification of proteins structures.

\parskip -6pt
\subsection*{SUPERFAMILY}
 SUPERFAMILY \cite{SUPERFAMILY-Gough01} was initially designed to provide protein domain assignments at the SCOP superfamily level. 
 The most recent release of SUPERFAMILY \cite{SUPERFAMILY-Wilson07} added the assignment of family level as well. SUPERFAMILY deploys a library of profile hidden Markov models (HMMs) that represent all proteins of known structure.  The HMM library is implemented based on Sequence Alignment and Modeling (SAM) HMM \cite{HMM-Wistrand05} package for sequence comparison. SUPERFAMILY's input is a protein structure. However, no domain boundary detection method is used. Instead, every model is scored (using a Viterbi algorithm) across the whole sequence detecting any occurrences of a domain belonging to the superfamily which the model represents. A specific heuristic strategy is used to select the regions of the sequence that correspond to a domain and match with a superfamily. In its latest release, SUPERFAMILY uses 10,894 models to represent the 1,539 superfamilies in SCOP 1.69. 
 The model and structural assignments are available from a public web server at http://supfam.org. 
To the best of our knowledge, there is no comparison of the accuracy of SUPERFAMILY with other existing methods in the literature. 

\subsection*{F2CS (CO)}
 Getz et al. \cite{byFSSP-Getz02} present a method, Classification by Optimization (CO), to automatically assign SCOP fold level and CATH topology level labels to input protein domains. This method is based on the FSSP (Fold classification based on Structure-Structure alignment of proteins) database \cite{FSSP} which uses DALI (Distance ALignment algorithm), a fully automated structure comparison algorithm, to calculate a pairwise structural similarity (the S-score) between protein chains. FSSP computes statistically meaningful Z-scores by shifting and rescaling S-scores, and uses Z-scores of all pairs of structures and a hierarchical clustering algorithm to generate a fold tree. CO is an optimization procedure that finds the assignment of minimal cost, where cost is defined in terms of Z-scores. The method is available online as a prediction server, F2CS \cite{F2CS-Getz04}, at {\small http://www.weizmann.ac.il/physics/complex/compphys/f2cs/}. The accuracy of CO method at predicting both SCOP fold level and CATH topology level was reported to be 93\%. Refer to \cite{byFSSP-Getz02} for details of their evaluation methodology.

\subsection*{SGM}
 Scaled Gauss Metric (SGM) \cite{SGM-Rogen03} is a measure of similarity of protein shapes based on Gauss integrals. In this measure, each domain is mapped into a point in $\Re^{30}$ (a 30-dimensional vector) and then the distance between two domains is defined to be the usual Euclidean distance between the points. Rogen and Fain \cite{SGM-Rogen03} show that SGM is a proper pseudo-metric satisfying, for example, triangle inequality on CATH (version 2.4) unlike previously proposed measures such as RMSD. They discovered that under this metric, protein structures naturally separate into fold clusters. Therefore, SGM is used to construct an automatic classification procedure for CATH database. The method is very fast since it does not need alignment of structure or all-pair comparison. It assigns 95.51\% of the input domains to proper CATH2.4 hierarchy (class, architecture, topology and homologous superfamily).

\subsection*{SCOPmap}
 SCOPmap \cite{SCOPMAP-Cheek04} has been developed to automatically map domains in protein structures to the SCOP database at the superfamily level. It also performs assignments at the SCOP fold level when confident superfamily level assignments cannot be made. Apart from finding appropriate SCOP superfamily for domains within newly solved proteins, SCOPmap can be used to find new links in SCOP by identifying potential evolutionary relationships between existing SCOP families. The general strategy used by this algorithm is to combine several existing sequence and structure comparison tools applied to a query protein of known structure to find the homologs already classified in SCOP and assign class labels based on the labels of those homologs. The tools used in SCOPmap include the gapped BLAST \cite{GPBLAST}, RPS-BLAST \cite{RPSBLAST}, PSI-BLAST \cite{GPBLAST},
COMPASS \cite{COMPASS}, MAMMOTH \cite{MAMMOTH}, and DaliLite \cite{DALILite}.
 This strategy is not limited to SCOP and can be used with other existing classification schemes as well. When applied to SCOP database, SCOPmap performs with roughly 95\% accuracy, i.e., for $\sim$95\% of the inputs, it predicts the superfamily correctly or no assignment is made as appropriate. SCOPmap performs better than SUPERFAMILY both in overall accuracy and in the detection of domain boundries.

\subsection*{DTree}

 A method for automatically classifying protein structures into SCOP classes based on decision trees, which we call DTree, is presented by \c{C}amoglu et al. \cite{DTree-Cam05}. In this approach, the decision of assigning a class label to an input domain is made by combining the decisions of multiple classifiers using a consensus of committee (or an ensemble) classifier. 
 Specifically, they use two sequence classifiers and three classifiers based on structure. Sequence classifiers are SUPERFAMILY \cite{SUPERFAMILY-Gough01} and a classifier based on PSI-BLAST \cite{GPBLAST}. The structure classifiers include a classifier based on CE algorithm for structural alignment \cite{CE}, DALI structure-similarity comparison tool \cite{DALI} and a classifier based on Vast \cite{vast} algorithm for identifying remote homologies.  
 Given an input protein domain, first the method determines whether the domain belongs to an existing category (family, superfamily, fold) in the SCOP hierarchy. For those that are predicted as members of current categories, the consensus classifier computes their family-, superfamily-, and fold-level classification. The authors present an elaborate comparison of their method with the individual classifiers and at different levels of classification. In summary, they show that their method based on decision trees achieves error rates that are 3-12 times less than the individual classifiers' error rates at the family level, 1.5-4.5 times less at the superfamily level, and 1.1-2.4 times less at fold level.

\subsection*{ProtClass}
 ProtClass (Protein Classification) \cite{ProtClass-Aung05} is an automatic structure classification method which does not require detailed structural alignment or binary classifications. The classifier in this method is built by using an existing classification database as the training data. The scheme presented in \cite{ProtClass-Aung05} is implemented and evaluated based on SCOP fold-level labels, although the authors state that other levels and classification databases can also be used. The classifier is trained by encoding the protein structure in each class into their concise formats and extracting some important pieces of information from each distinct class. In this way, prior knowledge and expert human judgement is used for automatically predicting the class labels for unseen structures. When classifying a new structure, again it is represented in its concise format, filtering and a nearest neighbour search is performed in order to efficiently and effectively predict the possible class label(s) for it. Experiments for evaluation of ProtClass are performed on using a 10-fold cross validation strategy on a relatively small dataset consisting of 600 proteins. The method is compared with the authors' previous work, SGM and a classification based on DALI \cite{DALI}. The results of their experiments show that ProtClass is slightly better than SGM in accuracy and much faster, but less accurate comparing with the classification based on DALI although extremely faster.

\subsection*{proCC} \parskip 0pt
 Kim and Patel \cite{proCC-Kim06} present proCC, a unified framework for structure classification and identification of novel protein structures. 
 This framework consists of three components. Given an unclassified query domain, first a structure comparison component employs an efficient index-based method to quickly find domains with similar structures. The comparison is performed using secondary structure elements (SSEs), which is more efficient than using atomic coordinates of C$_{\alpha}$ atoms. The scoring function used is similar to that of DALI \cite{DALI} for tertiary structure comparison, but uses SSEs as basic unit of comparison rather than individual residues. This makes this scoring method computationally much faster than DALI. The second component of proCC is a classification component that assigns the query to an existing class label or marks it as {\em unclassified} based on the results of the structure comparison component. The classification of a query is performed using top $k$ structure neighbours in the database and a support vector machine (SVM) trained using training data (known class labels in SCOP or CATH). Finally, a clustering component takes all domains marked as unclassified and detects potentially novel folds by clustering them using a graph-clustering method on a weighted graph constructed from all unclassified structures. In this graph, nodes represent the unclassified structures. Two nodes are connected if their similarity score is above a certain threshold and the similarity score is the weight of that edge.
 
 Comparing with other automatic classification methods, a major feature of proCC  is that not only it is capable of accurately classifying new domains into {\em existing} classes, it can effectively identify {\em new} classes.
 Emphasis in the evaluation of proCC is on identification of novel domains. For example, for comparison with SGM, the parameters of SGM are set to produce the best accuracy for new class detection, and then it is shown that proCC outperforms SGM in classification precision and error rates. In the evaluation on SCOP version 1.69, proCC correctly classifies 86.0\%, 87.7\%, and 90.5\% of new domains at familiy, superfamily, and fold levels. Comparing with SGM (when its parameters are set to produce the best accuracy for detecting new classes), proCC performs 15-19\% more accurate in the family, superfamily and fold levels of SCOP , and is slightly more accurate than SGM in classifying CATH. In the detection of new classes, SGM could perform equally well. Comparing with SCOPmap, proCC is marginally better in overal precision but is about 20\% more accurate in detecting novel structures. Note that unlike SCOPmap, proCC has to use another tool for domain boundary detection.

\parskip -4pt
\subsection*{fastSCOP} 
 The fastSCOP \cite{fastSCOP-Tung07} server is the most recent proposal for automated classification of protein structures. The fastSCOP is a web server that quickly recognizes protein structural domains and SCOP superfamilies of a query protein structure. The server uses 3D-BLAST \cite{3DBLAST} (a method developed by the same authors, for quickly finding similar structures) to scan quickly a large classification database, namely latest release of SCOP. The top 10 hit domains which have different superfamily classifications are obtained. MAMMOTH \cite{MAMMOTH}, a detailed structural alignment tool, is adopted to align these structures to refine domain boundaries and identify SCOP superfamilies. The classification accuracy of this server on 586 query structures (including 464 single-domain and 122 multi-domain proteins that are in SCOP 1.69 but not in SCOP 1.67) on SCOP 1.67 is $\sim$98\% with $\sim$5 seconds average execution time on a modest personal computer. The fastSCOP web server is available at http://fastscop.life.nctu.edu.tw.
 


\section*{Outlook}

 Automated classification of protein structures is an attractive research topic with many interesting directions for future work. In this section, we discuss some open problems as well as interesting ideas for further improvement in the accuracy and efficiency of the classification procedure. One important point to have in mind is that although many methods have been developed, a fully automatic high-quality classification database or a semi-automatic database that is updated frequently enough is yet to be developed. This remains the main challenge for computational biologists active in this area. An important step to be taken for manual and semi-automatic methods, SCOP and CATH, is to utilize several new techniques discussed in this paper to speed up their classification process and keep their high-quality classification databases more up-to-date without losing the quality of their classification. \\

\noindent \textbf{Possible improvements in existing methods:} All the methods discussed in this paper have still open problems and possibility for further improvements. \textbf{SUPERFAMILY} uses a specific domain boundary detection mechanism as a part of its methodology. More effective boundary detection techniques can replace the boundary detection mechanism in SUPERFAMILY. \textbf{F2CS(CO)} uses a simple clustering algorithm (average-linkage hierarchical clustering algorithm) to generate its fold tree. More accurate and efficient hierarchical clustering algorithms can improve the effectiveness of this method. The \textbf{SGM} method opens up lots of interesting future directions by the novel measure of similarity of structures proposed based on Gauss integrals.
SGM method is initially proposed to accurately classify domains into CATH hierarchy. The evaluation in \cite{proCC-Kim06} shows SGM does not perform equally well on SCOP. SGM method can be extended for prediction of SCOP hierarchy. For this, the classification algorithm should be made more elaborate by incorporating cluster-specific information. SGM method can also be combined with a domain boundary detection method and be extended to predict domain boundaries of input structures as well. The \textbf{SCOPmap} method, as stated in the description of its methodology, can be extended for prediction of CATH hierarchy as well. An interesting feature of SCOPmap is that it uses several existing sequence and structure comparison tools as a part of its classification process. Therefore, new proposals for highly accurate and efficient sequence and structure comparison such as 3D-BLAST\cite{3DBLAST} can be added to this method to make it even more effective. \textbf{DTree} has also the feature of incorporating existing tools for building its classifier based on decision trees. Similarly, these tools can be replaced by recent more effective techniques. An obvious extension would be using fastSCOP instead of SUPERFAMILY in DTree to significantly improve the performance of the method. In \textbf{ProtClass}, the nearest-neighbour classification strategy can be replaced by a more effective classification technique. The flexible framework of \textbf{proCC} can also be extended in several ways. It can be combined with a more effective boundary detection method such as recently proposed CATHEDRAL \cite{CATHEDRAL-Redfern07} to further improve its effectiveness comparing with SCOPmap. Its clustering method can also be enhanced with more effective graph clustering algorithms to enhance its accuracy in identification of novel folds. The \textbf{fastSCOP} server is very accurate and highly efficient but is only capable of classifying at SCOP superfamily level. Extending fastSCOP's approach for CATH classification and other levels of SCOP would be an enticing direction for future work.
\\

\noindent \textbf{Automated clustering of structures:}
Instead of {\em classifying} input structures into known classes of existing databases, structures can be {\em clustered} into groups of related structures. A very recent work \cite{STRALCP-Zemla07} presents STRALCP, a method for STRuctural ALignment-based Clustering of Protein structures. The goal of this method is to automatically identify structurally conserved regions for a given set of protein structures and use them to cluster results similar to those that would be obtained by manual inspection (e.g., SCOP curators). STRALCP uses a specific structure alignment algorithm as a part of its clustering algorithm. Several other alignment and clustering algorithms exist that could be investigated for generating clusters similar to SCOP or CATH, or results that could be used in the automated and semi-automatic classifications techniques. \\

\noindent \textbf{Classification based on information bottleneck method:}.
 Recently, there has been an increasing interest in methods based on information theory concepts for clustering and classification techniques in the information retrieval and databases community \cite{thesisIB, LIMBO}. 
One such method is called information bottleneck (IB). In a simple IB algorithm, clustering is performed by first assuming that
each record is a separate cluster and then iteratively merging the clusters $n - k$ times to reduce the number of clusters to $k$. In each iteration, two clusters are chosen to be merged so that the amount of \textit{information loss} as a result of merging the
clusters is minimum. Information loss is given by a specific formula derived from information theory concepts. This formula could be adopted for clustering and classification of structures based on a structure comparison measure such as Z-Scores of DALI \cite{DALI}.
\\

\noindent \textbf{A hybrid method:} As described in the above sections and also shown in Tables \ref{tab1} and \ref{tab2}, each classification method is designed and performs better for a specific output. For example, fastSCOP is only capable of prediction of SCOP superfamily, although it is highly accurate and efficient. The accurate results of fastSCOP in superfamily level can be combined with results of another method such as proCC that can classify other levels, in order to predict full classification hierarchy of an input structure. This is in particular useful in methods like DTree and SCOPmap that use other tools as part of their classification process and also proCC and SCOPmap that make the decision based on top $k$ similar structures in an existing database. Using this technique, irrelevant results from the top $k$ structures can be pruned.
\\

\noindent  \textbf{Other supervised learning methods:}
 Several methods described in this paper are based on a supervised learning technique. A classifier learns the classification rules from the training data and reapplies these rules to classify new unseen inputs. There exists several classification methods in the machine learning literature including nearest-neighbor search (NN), support vector machines (SVM), neural networks, and Bayesian networks, some of which were used as a part of the methods presented in this paper. Adopting novel classification methods for existing classification methods or in a novel framework could be an interesting future work. Hidden Markov models (HHMs) are used for sequence comparison as in SUPERFAMILY method. A similar method based on HMMs but on SSEs (as in proCC) or other representations of protein structures (as in SGM or ProtClass) is one possible novel classification technique.

\section*{Conclusion}

 We presented a survey of recent proposals for automated classification of protein structures. We provided an overall comparison of the methodology of different methods and briefly explained important features of each technique. Automated classification of protein structures is still an attractive research topic with a lot of interesting possibilities for further improvements in the accuracy and efficiency. Recent developments in detection of domain boundaries \cite{CATHEDRAL-Redfern07} can be used to improve the performance of a flexible framework such as proCC \cite{proCC-Kim06}. More advanced machine learning techniques can also be used for training an accurate classifier. Another interesting direction for future work is adopting ideas from information theory field to provide an accurate and efficient algorithm for classification of protein structures. Having developed several accurate and efficient classification methods, the manual and semi-automatic classification databases are now able to utilize these methods to keep their high-quality classification more up-to-date.

\nocite{SUPERFAMILY-Gough01, byFSSP-Getz02, SGM-Rogen03, SCOPMAP-Cheek04, DTree-Cam05, ProtClass-Aung05,  proCC-Kim06, fastSCOP-Tung07, STRALCP-Zemla07, CATHEDRAL-Redfern07, PDB}


\newpage

\begin{thebibliography}{10}

\bibitem{FSSP}
Liisa Holm and Chris Sander.
\newblock {Dali/FSSP} classification of three-dimensional protein folds.
\newblock {\em Nucleic Acids Research}, 25(1):231--234, 1997.

\bibitem{CATH}
C.~A. Orengo, A.~D. Michie, S.~Jones, D.~T. Jones, M.~B. Swindells, and J.~M.
  Thornton.
\newblock {CATH}--a hierarchic classification of protein domain structures.
\newblock {\em Structure}, 5(8):1093--1108, August 1997.

\bibitem{SCOP}
L.~Lo~Conte, B.~Ailey, T.~J. Hubbard, S.~E. Brenner, A.~G. Murzin, and
  C.~Chothia.
\newblock {SCOP}: a structural classification of proteins database.
\newblock {\em Nucleic Acids Research}, 28(1):257--259, January 2000.

\bibitem{HOMSTRAD}
K.~Mizuguchi, C.~M. Deane, T.~L. Blundell, and J.~P. Overington.
\newblock {HOMSTRAD}: a database of protein structure alignments for homologous
  families.
\newblock {\em Protein Sci}, 7(11):2469--2471, November 1998.

\bibitem{MMDB}
J.~F. Gibrat, T.~Madej, and S.~H. Bryant.
\newblock Surprising similarities in structure comparison.
\newblock {\em Current Opinion in Structural Biology}, 6(3):377--85, 1996.

\bibitem{3Dee}
Asim~S. Siddiqui, Uwe Dengler, and Geoffrey~J. Barton.
\newblock {3Dee}: a database of protein structural domains.
\newblock {\em Bioinformatics}, 17(1):200--201, 2001.

\bibitem{PDB}
H.~M. Berman, J.~Westbrook, Z.~Feng, G.~Gilliland, T.~N. Bhat, H.~Weissig,
  I.~N. Shindyalov, and P.~E. Bourne.
\newblock The protein data bank.
\newblock {\em Nucleic Acids Res}, 28(1):235--242, January 2000.

\bibitem{SUPERFAMILY-Gough01}
J.~Gough, K.~Karplus, R.~Hughey, and C.~Chothia.
\newblock Assignment of homology to genome sequences using a library of hidden
  markov models that represent all proteins of known structure.
\newblock {\em J Mol Biol}, 313(4):903--919, November 2001.

\bibitem{SUPERFAMILY-Wilson07}
D~Wilson, M~Madera, C~Vogel, C~Chothia, and J~Gough.
\newblock The {SUPERFAMILY} database in 2007: families and functions.
\newblock {\em Nucleic Acids Res}, 35(Database issue):308--313, Jan 2007.

\bibitem{byFSSP-Getz02}
Gad Getz, Michele Vendruscolo, David Sachs, and Eytan Domany.
\newblock Automated assignment of {SCOP} and {CATH} protein structure
  classifications from {FSSP} scores.
\newblock {\em Proteins: Structure, Function, and Genetics}, 46(4):405--415,
  January 2002.

\bibitem{F2CS-Getz04}
Gad Getz, Alina Starovolsky, and Eytan Domany.
\newblock {F2CS}: {FSSP} to {CATH} and {SCOP} prediction server.
\newblock {\em Bioinformatics}, 20(13):2150--2152, 2004.

\bibitem{SGM-Rogen03}
P.~Rogen and B.~Fain.
\newblock Automatic classification of protein structure by using {Gauss}
  integrals.
\newblock {\em Proc Natl Acad Sci U S A}, 100(1):119--124, January 2003.

\bibitem{SCOPMAP-Cheek04}
S.~Cheek, Y.~Qi, S.~S. Krishna, L.~N. Kinch, and N.~V. Grishin.
\newblock {SCOPmap}: automated assignment of protein structures to evolutionary
  superfamilies.
\newblock {\em BMC Bioinformatics}, 5(1), December 2004.

\bibitem{DTree-Cam05}
O.~Camo\u{g}lu, T.~Can, A.~K. Singh, and Y.~F. Wang.
\newblock Decision tree based information integration for automated protein
  classification.
\newblock {\em J Bioinform Comput Biol}, 3(3):717--742, June 2005.

\bibitem{ProtClass-Aung05}
Z.~Aung and K.~L. Tan.
\newblock Automatic {3D} protein structure classification without structural
  alignment.
\newblock {\em J Comput Biol}, 12(9):1221--1241, November 2005.

\bibitem{proCC-Kim06}
You~J. Kim and Jignesh~M. Patel.
\newblock A framework for protein structure classification and identification
  of novel protein structures.
\newblock {\em BMC Bioinformatics}, 7:456+, October 2006.

\bibitem{STRALCP-Zemla07}
A.~Zemla, B.~Geisbrecht, J.~Smith, M.~Lam, B.~Kirkpatrick, M.~Wagner,
  T.~Slezak, and CE. Zhou.
\newblock {STRALCP} structure alignment-based clustering of proteins.
\newblock {\em Nucleic Acids Res}, 35, 2007.

\bibitem{HMM-Wistrand05}
M.~Wistrand and E.~L. Sonnhammer.
\newblock Improved profile {HMM} performance by assessment of critical
  algorithmic features in {SAM} and {HMMER}.
\newblock {\em BMC Bioinformatics}, 6(1), April 2005.

\bibitem{GPBLAST}
S.~F. Altschul, T.~L. Madden, A.~A. Sch\"{a}ffer, J.~Zhang, Z.~Zhang,
  W.~Miller, and D.~J. Lipman.
\newblock {Gapped BLAST and PSI-BLAST}: a new generation of protein database
  search programs.
\newblock {\em Nucleic Acids Res}, 25(17):3389--3402, September 1997.

\bibitem{RPSBLAST}
A.~Marchler-Bauer, J.~B. Anderson, C.~DeWeese-Scott, N.~D. Fedorova, L.~Y.
  Geer, S.~He, D.~I. Hurwitz, J.~D. Jackson, A.~R. Jacobs, C.~J. Lanczycki,
  C.~A. Liebert, C.~Liu, T.~Madej, G.~H. Marchler, R.~Mazumder, A.~N.
  Nikolskaya, A.~R. Panchenko, B.~S. Rao, B.~A. Shoemaker, V.~Simonyan, J.~S.
  Song, P.~A. Thiessen, S.~Vasudevan, Y.~Wang, R.~A. Yamashita, J.~J. Yin, and
  S.~H. Bryant.
\newblock {CDD}: a curated entrez database of conserved domain alignments.
\newblock {\em Nucleic Acids Res}, 31(1):383--387, January 2003.

\bibitem{COMPASS}
R.~Sadreyev and N.~Grishin.
\newblock {COMPASS}: a tool for comparison of multiple protein alignments with
  assessment of statistical significance.
\newblock {\em J Mol Biol}, 326(1):317--336, February 2003.

\bibitem{MAMMOTH}
Angel~R. Ortiz, Charlie~E. Strauss, and Osvaldo Olmea.
\newblock {MAMMOTH} (matching molecular models obtained from theory): An
  automated method for model comparison.
\newblock {\em Protein Sci}, 11(11):2606--2621, November 2002.

\bibitem{DALILite}
Liisa Holm and Jong Park.
\newblock {DaliLite} workbench for protein structure comparison.
\newblock {\em Bioinformatics}, 16(6):566--567, 2000.

\bibitem{CE}
I.~N. Shindyalov and P.~E. Bourne.
\newblock Protein structure alignment by incremental combinatorial extension
  ({CE}) of the optimal path.
\newblock {\em Protein Eng}, 11(9):739--747, September 1998.

\bibitem{DALI}
L.~Holm and C.~Sander.
\newblock Protein structure comparison by alignment of distance matrices.
\newblock {\em J Mol Biol}, 233(1):123--138, September 1993.

\bibitem{vast}
T.~Madej, J.~F. Gibrat, and S.~H. Bryant.
\newblock Threading a database of protein cores.
\newblock {\em Proteins}, 23(3):356--369, November 1995.

\bibitem{fastSCOP-Tung07}
Chi-Hua Tung and Jinn-Moon Yang.
\newblock {fastSCOP}: a fast web server for recognizing protein structural
  domains and scop superfamilies.
\newblock {\em Nucleic Acids Res}, 35, 2007.

\bibitem{3DBLAST}
J.~M. Yang and C.~H. Tung.
\newblock Protein structure database search and evolutionary classification.
\newblock {\em Nucleic Acids Res}, 34(13):3646--3659, 2006.

\bibitem{CATHEDRAL-Redfern07}
Oliver~C. Redfern, Andrew Harrison, Tim Dallman, Frances~M. Pearl, and
  Christine~A. Orengo.
\newblock {CATHEDRAL}: A fast and effective algorithm to predict folds and
  domain boundaries from multidomain protein structures.
\newblock {\em PLoS Computational Biology}, 3(11):e232+, November 2007.

\bibitem{thesisIB}
N.~Slonim.
\newblock The information bottleneck: Theory and applications.
\newblock PhD Thesis.

\bibitem{LIMBO}
Periklis Andritsos, Panayiotis Tsaparas, Ren{\'e}e~J. Miller, and Kenneth~C.
  Sevcik.
\newblock {LIMBO}: Scalable clustering of categorical data.
\newblock In {\em EDBT}, pages 123--146, 2004.

\end{thebibliography}
\end{document}